\documentclass{IEEEtran}[10pt,twocolumn,journal]%peerreview]
\usepackage[noadjust]{cite}
\usepackage{times,amsmath,graphicx,epsfig,amssymb, mathtools} %,amssymb,eucal
\usepackage[T1]{fontenc}
\newcommand{\diag} {\mathop{\rm diag}}

\newcommand{\T} {\mathop{\rm \textsf{T}}}
\newcommand{\F} {\mathop{\rm \textsf{F}}}
\newcommand{\Ff} {\mathop{\rm \mathcal{F}}}
\newcommand{\I} {\mathop{\rm {\boldsymbol{I}}}}
\newcommand{\0} {\mathop{\rm \boldsymbol{0}}}

\newcommand{\jj} {\mathop{\rm {\mathsf{j}}}}
\newcommand{\D} {\mathop{\rm {\mathcal{D}}}}

\newcommand{\Ss} {\mathop{\rm {\mathcal{S}}}}

\newcommand{\HH} {\mathop{\rm {\mathcal{H}}}}

\newcommand{\Cc} {\mathop{\rm {\mathcal{C}}}}
\newcommand{\CN} {\mathop{\rm {\mathcal{CN}}}}

\newcommand{\Vect}[1] {\boldsymbol{#1}}

\newtheorem{definition}{Definition}%[section]
\newtheorem{lemma}{Lemma}%[section]
\newtheorem{theorem}{Theorem}%[section]
\begin{document}

\title{Four-Group Decodable Space-Time Block Codes}
\author{D\~{u}ng~Ng\d{o}c~Ðào, \IEEEmembership{Member,~IEEE}, Chau Yuen, \IEEEmembership{Member,~IEEE}, Chintha~Tellambura, \IEEEmembership{Senior Member,~IEEE}, Yong Liang Guan, \IEEEmembership{Member,~IEEE}, and Tjeng Thiang Tjhung, \IEEEmembership{Senior Member,~IEEE}% <-this % stops a space
\thanks{Manuscript received  November 7, 2006; revised February 23, 2007, and May 7, 2007. The work of D. N. Ðào and C. Tellambura was supported by The National Sciences and Engineering Research Council
(NSERC) and Alberta Informatics Circle of Research Excellence
(iCORE), Canada. The editor coordinating the review of this paper and approving it for publication was Dr. Franz Hlawatsch.}
\thanks{D. N. \DH \`{a}o was with Department of
Electrical and Computer Engineering, University of Alberta,
Edmonton, Alberta T6G 2V4, Canada. He is now with Department of
Electrical and Computer Engineering, McGill University, Montr\'{e}al, Qu\'{e}bec, H3A 2A7, Canada. (e-mail: ngoc.dao@mail.mcgill.ca)}
\thanks{C. Yuen and T. T. Tjhung are with Institute for Infocomm Research, 21 Heng Mui Keng Terrace, Singapore 119613. e-mail: \{cyuen,  tjhungtt\}@i2r.a-star.edu.sg.}
\thanks{C. Tellambura is with Department of Electrical and Computer Engineering, University of Alberta, Edmonton, Alberta T6G 2V4, Canada. (e-mail: chintha@ece.ualberta.ca)}
\thanks{Y. L. Guan is with the School of Electrical and Electronic Engineering, Nanyang Technological University, S1-B1c-108, Nanyang Avenue, Singapore, 639798. e-mail: eylguan@ntu.edu.sg}}

\markboth{IEEE Transactions on Signal Processing,~Vol.~X,
No.~X,~X~200X}{\DH \`{a}o \MakeLowercase{\textit{et al.}}:
Four-Group Decodable Space-Time Block Codes}

\maketitle

\begin{abstract}
Two new rate-one full-diversity space-time block codes (STBC) are
proposed. They are characterized by the \emph{lowest decoding
complexity} among the known rate-one STBC, arising due to the
complete separability of the transmitted symbols into four groups
for maximum likelihood detection.  The first and the second codes
are delay-optimal if the number of transmit antennas is a power of 2
and even, respectively. The exact pair-wise error probability is
derived to allow for the performance optimization of the two codes.
Compared with existing low-decoding complexity STBC, the two new
codes offer several advantages such as higher code rate, lower
encoding/decoding delay and complexity, lower peak-to-average power
ratio, and better performance.
\end{abstract}

\begin{keywords}
Orthogonal designs, performance analysis, quasi-orthogonal space-time block codes, space-time block codes.
\end{keywords}

\IEEEpeerreviewmaketitle

\section{Introduction}

Space-time block codes (STBC\footnote{The term "STBC" stands for
space-time block code/codes/coding, depending on  the context.}) have
been extensively studied since they  exploit the diversity and/or
the capacity of multiple-input multiple-output (MIMO) channels.
Among various STBC, orthogonal STBC (OSTBC) \cite{alamouti98,
tarokh99a, liang03a} offer  the minimum decoding complexity and full
diversity. However, they have low code rates when the number of
transmit (Tx) antennas is more than 2 \cite{liang03a}. The rate of
one symbol per channel use (pcu) only exists for 2 Tx antennas  and
the rate approaches 1/2 for a large number of Tx antennas
\cite{alamouti98, tarokh99a, liang03a}.

To improve the low rate of OSTBC, several quasi-orthogonal STBC
(QSTBC) have been proposed (see \cite{jafar01a, tirk00, su04b,
bad04b} and references therein). They allow joint maximum likelihood
(ML) decoding of pairs of complex symbols. However, the rate-one
QSTBC exist for 4 Tx antennas only and the code rate is smaller than
1 for more than 4 Tx antennas. Several rate-one STBC have been
proposed (e.g. \cite{sha03b, sez04b, dao05a}), in which the
transmitted symbols can be completely separated into two groups for
ML detection. However, for more than 4 Tx antennas, the decoding
complexity of  the rate-one STBC in \cite{sha03b, sez04b, dao05a}
increases  significantly compared with OSTBC and QSTBC.

In this paper, we propose two new rate-one STBC for any number of Tx
antennas. Compared  with the existing rate-one STBC, our new codes
have lowest decoding complexity since the transmitted symbols can be
decoupled into 4 groups (4Gp) for ML detection. The first code is
called 4Gp-QSTBC. The second code is derived from semi-orthogonal
algebraic space-time (SAST) codes \cite{dao05a} and thus called
4Gp-SAST codes. The first and the second codes are delay-optimal
when the number of Tx antennas is a power of 2 and even,
respectively. The equivalent transmit-receive signals are derived so
that sphere decoders \cite{damen03a} can be applied for data
detection. To achieve full-diversity, signal rotations are required
for the two codes. The exact pair-wise error probability (PEP) of
the two codes is derived to optimize the signal rotations.

We compare the main parameters of our new codes  and several
existing STBC for 6 and 8 Tx antennas in Table \ref{t1}. Clearly,
the new codes offer several distinct advantages such as higher code
rate, low decoding complexity, and lower encoding/decoding delay.
The two new codes also have lower peak-to-average power ratio (PAPR)
than OSTBC, QSTBC, and minimum decoding complexity (MDC) QSTBC
\cite{yuen05b}. Moreover, simulation results show that  our new
codes also yield significant SNR gains compared with the existing
codes.

\begin{table}[b]
\renewcommand{\arraystretch}{1.2}
\centering
  \caption{Comparison of Several Low Complexity STBC for 6 and 8 Antennas. The Numbers in the Parentheses Indicate the Codes' Parameters for 8 Tx Antennas.}\label{t1}
  \begin{tabular}{|c|c|c|c|c|}
    \hline
    \hline
    % after \\: \hline or \cline{col1-col2} \cline{col3-col4} ...Constellation
       Codes  & Maximal rate & Delay & Real symbol decoding  \\
    \hline
    \hline
       OSTBC \cite{liang03a, kan05a}  &  2/3 (5/8) & 30 (56) & 1 or 2 (1 or 2)  \\
    \hline
       CIOD \cite{khan06a}  &6/7 (4/5) & 14 (50) &  2 (2) \\
    \hline
       MDC-QSTBC \cite{yuen05b}  & 3/4 (3/4) & 8 (8) & 2 (2)  \\
    \hline
       QSTBC \cite{su04b} & 3/4 (3/4) & 8 (8) & 4 (4)  \\
    \hline
       2Gp-QSTBC \cite{sha03b}  & 1 (1) & 8 (8) & 8 (8) \\
    \hline
       SAST \cite{dao05a} & 1 (1) & 6 (8) & 6 (8) \\
    \hline
       \textbf{4Gp-QSTBC} (new) & \textbf{1} (\textbf{1}) & \textbf{8} (\textbf{8}) & \textbf{4} (\textbf{4})  \\
    \hline
       \textbf{4Gp-SAST} (new) & \textbf{1} (\textbf{1}) & \textbf{6} (\textbf{8}) & \textbf{3} (\textbf{4})  \\
    \hline
    \hline
  \end{tabular}
\end{table}

\emph{Notation}: Superscripts $^{\T}$, $^*$, and $^\dag$ denote matrix transpose,
conjugate, and transpose conjugate, respectively. The identity and
all-zero  square matrices of proper size are denoted by ${\I}$ and
$\0$. The diagonal matrix with elements of vector $\Vect{x}$ on the
main diagonal is denoted by $\diag(\Vect{x})$. $\|{X}\|_{\F}$ stands
for the Frobenius norm of matrix $X$ and $\otimes$ denotes Kronecker
product \cite{davis79}. A mean-$m$ and
variance-$\sigma^2$ circularly complex Gaussian random variable is
written by $\mathcal{CN}(m, \sigma^2)$. $\Re(X)$ and $\Im(X)$ denote
the real and imaginary parts of $X$, respectively.

\section{System Model and Preliminaries}

\subsection{System Model}

We consider data transmission over a MIMO quasi-static Rayleigh flat
fading channel with $M$ Tx and $N$ receive (Rx) antennas
\cite{hass02}. The channel gain $h_{mn}$ $(m = 1, 2, \ldots, M; n =
1, 2, \ldots, N)$ between the $(m, n$)-th Tx-Rx antenna pair is
assumed $\mathcal{CN}(0, 1)$ and remains constant over $T$ time
slots. We assume no spatial correlation at either Tx or Rx array.
The receiver, but not the transmitter, completely knows the channel
gains.

A $T \times M$ STBC
can be represented in a general dispersion form \cite{hass02} as
follows:
\begin{equation}\label{e4}
    X = \sum_{k=1}^{K}{\left(a_k A_k + b_k B_k \right)} %\qquad ({\jj}^2 = -1),
\end{equation}
\noindent where $A_k$ and $B_k$, ($k = 1, 2, \cdots, K$) are $T
\times M$ constant matrices, commonly called dispersion matrices;
$a_k$ and $b_k$ are the real and imaginary parts of the symbol
$s_k$. We can use an equivalent  form of STBC as
\begin{equation}\label{e5}
    X = \sum_{l=1}^{L}{c_l C_l}
\end{equation}
where $L$ is the number (not necessarily  even) of transmitted
symbols, $c_l$ are real-value transmitted symbols, $C_l$ are
dispersion matrices. The average energy of code matrices is
constrained such that $\mathcal{{E}}_{\mathcal{X}} =
\mathbb{E}[\|{X}\|^2_{\F}] = T$.

The received signals $y_{tn}$ of the $n$th antenna at time $t$ can
be arranged in a matrix $Y$ of size $T \times N$. Thus, one can
represent the Tx-Rx signal relation as  \cite{tarokh98a, hass02}
\begin{equation}\label{e3}
    Y = \sqrt{\rho}XH + Z
\end{equation}
where $H = [h_{mn}]$ is the channel matrix; $Z = [z_{tn}]$ is the noise matrix of size $T \times N$, its elements $z_{tn}$ are independently, identically distributed (i.i.d.)
$\mathcal{CN}(0, 1)$. The Tx power is scaled by ${\rho}$ so
that the average signal-to-noise ratio (SNR) at each Rx antenna
is $\rho$, independent of the number of Tx antennas.

Let the data vector be $\Vect{c} = \begin{bmatrix} c_1 & c_2 &
\ldots c_L \end{bmatrix}^{\T}$.  The ML decoding of STBC is to find
the solution $\hat{\Vect{c}}$ so that:
\begin{equation}\label{e6}
    \hat{\Vect{c}} = \arg \min_{\Vect{c}} \|Y - XH\|_{\F}^2\,.
\end{equation}

\subsection{Algebraic Constraints of QSTBC}

The key idea of  QSTBC is to divide the $L$ (real) transmitted
symbols embedded in a  code matrix into $\Gamma$ groups, so that the
ML detection of the  transmitted symbol  vector can be decoupled
into $\Gamma$ sub-metrics, each metric involves the symbols of only
one group \cite{su04b,  sha03b, yuen05c, dao05a}. We provide a
definition of STBC with this  feature to unify the notation in this
paper as follows.

\begin{definition} \label{d1}
    \emph{A STBC is said to be $\Gamma$-group decodable STBC if the ML  decoding metric \eqref{e6} can be decoupled into a linear sum of $\Gamma$ independent submetrics, each submetric consists of the symbols from only one group. The $\Gamma$-group decodable STBC is denoted by $\Gamma$Gp-STBC for short.}
\end{definition}

In the most general case, we assume that there are $\Gamma$ groups;
each group  is denoted by $\Omega_i$ $(i = 1, 2, \ldots, \Gamma)$
and has $L_i$ symbols. Thus $L = \sum_{i=1}^{\Gamma} L_i$. Let
$\Theta_i$ be the set of indexes of symbols in the group $\Omega_i$.

Yuen \emph{et al.} \cite[Theorem 1]{yuen05c} have  shown a
sufficient condition for a STBC to be $\Gamma$-group decodable. In
fact, this condition is also necessary. We will state these results
in the following theorem without proof for brevity.

\begin{theorem} \label{th1}
\emph{The necessary and sufficient conditions, so that a STBC is $\Gamma$-group decodable, are}
\begin{equation} \label{e7}
   C_p^\dag C_q + C_q^\dag C_p = \0 \quad \forall p \in \Theta_i, \forall q \in \Theta_j, i \neq j.        \end{equation}
\end{theorem}

Note that Theorem \ref{th1} covers \cite[Theorem 9]{khan06a}
(single-symbol decodable STBC) and can be shown similarly.

\section{Four-group Decodable STBC Derived from QSTBC}

\subsection{Encoding}

In this section, we will study the new 4Gp-QSTBC. As we will see
later,  the general form of STBC in \eqref{e4} is convenient for
studying 4Gp-QSTBC; hence Theorem \ref{th1} can be restated as follows.

\begin{lemma}[\cite{yuen05d}] \label{l1}
\emph{The necessary and sufficient conditions for a STBC in
\eqref{e4} to  become $\Gamma$-group decodable  are}: (a) $A_p^\dag
A_q + A_p^\dag A_q = \0$, (b) $B_p^\dag B_q + B_p^\dag B_q = \0$,
and (c) $A_p^\dag B_q + B_p^\dag A_q = \0$, $\forall p \in \Theta_i,
\forall q \in \Theta_j, 1 \leq i \neq j \leq \Gamma$.
\end{lemma}

We next consider another sufficient condition so that a STBC is
four-group decodable.

\begin{theorem} \label{th2}
\emph{Given a 4Gp-STBC for $M$ Tx antennas with code length
$T$ and $K$ sets of dispersion matrices $(A_k, B_k; 1 \leq k \leq
K)$, a 4Gp-STBC with code length $2T$ for $2M$ Tx antennas,
which consists of $2K$ sets of dispersion matrices denoted as
$(\bar{A}_i, \bar{B}_i), 1 \leq i \leq 2K$, can be constructed using
the following mapping rules:}
\begin{align} \label{e13}
    &\bar{A}_{2k-1} = \begin{bmatrix}
        A_k & \0  \\
        \0 & A_k  \\
      \end{bmatrix},
    \; \bar{A}_{2k} = \begin{bmatrix}
         B_k & \0  \\
        \0 &  B_k  \\
      \end{bmatrix}, \notag \\
     &\bar{B}_{2k-1} = \begin{bmatrix}
         \0 &  A_k  \\
          A_k & \0   \\
      \end{bmatrix},
  \; \bar{B}_{2k} = \begin{bmatrix}
         \0 & B_k  \\
         B_k & \0   \\
      \end{bmatrix}.
\end{align}
\end{theorem}

\begin{proof}
Theorem \ref{th2} can be proved by showing that  if the dispersion
matrices $(A_q, B_q)\, (1 \leq q \leq K)$ satisfy Lemma \ref{l1}
with  $(A_p, B_p)\, (1 \leq p \leq K)$ where $q \notin \Theta_p$,
then the dispersion matrices $(\bar{A}_{2q-1}, \bar{B}_{2q-1},
\bar{A}_{2q}, \bar{B}_{2q})$ constructed from $(A_q, B_q)$ using
\eqref{e13} will satisfy Theorem \ref{th2} with  $(\bar{A}_{2p-1},
\bar{B}_{2p-1}, \bar{A}_{2p}, \bar{B}_{2p})$ constructed from $(A_p,
B_p)$ using \eqref{e13}. The detailed proof is omitted here, as the
steps are routine.
\end{proof}

The recursive construction of 4Gp-STBC specified in Theorem
\ref{th2} suggests that we can start with the MDC-QSTBC for 4
Tx antennas proposed in \cite{yuen05b} to construct 4Gp-STBC
for 8, 16  Tx antennas and so on, because MDC-QSTBC is one of
the STBC satisfying Lemma \ref{l1}; the resulting STBC is thus
called 4Gp-QSTBC. For practical interest, we will illustrate the
encoding process of 4Gp-QSTBC for 8 Tx antennas from the
MDC-QSTBC for 4 Tx antennas \cite{yuen05b}. The code matrix of
MDC-QSTBC for 4 Tx antennas is
\begin{align}  \label{e14} %\phantom{-}
    F_4 &= \begin{bmatrix*}[r]
    a_1 + \jj a_3 & a_2 + \jj a_4 & b_1 + \jj b_3 & b_2 + \jj b_4  \\
   -a_2 + \jj a_4 & a_1 - \jj a_3 &-b_2 + \jj b_4 & b_1 - \jj b_3  \\
    b_1 + \jj b_3 & b_2 + \jj b_4 & a_1 + \jj a_3 & a_2 + \jj a_4  \\
   -b_2 + \jj b_4 & b_1 - \jj b_3 &-a_2 + \jj a_4 & a_1 - \jj a_3  \\
      \end{bmatrix*}
\end{align}
where ${\jj}^2 = -1$.

The code matrix of 4Gp-QSTBC for 8 Tx
antennas from $F_4$ using mapping rules in \eqref{e13} is given below:
\begin{align} \label{e15}  %\phantom{-}
    F_8 &=  \left[\begin{array}{rrrrr}
    a_1 + \jj a_5 & a_3 + \jj a_7 & a_2 + \jj a_6 & a_4 + \jj a_8\\
   -a_3 + \jj a_7 & a_1 - \jj a_5 &-a_4 + \jj a_8 & a_2 - \jj a_6\\
    a_2 + \jj a_6 & a_4 + \jj a_8 & a_1 + \jj a_5 & a_3 + \jj a_7\\
   -a_4 + \jj a_8 & a_2 - \jj a_6 &-a_3 + \jj a_7 & a_1 - \jj a_5\\
    b_1 + \jj b_5 & b_3 + \jj b_7 & b_2 + \jj b_6 & b_4 + \jj b_8\\
   -b_3 + \jj b_7 & b_1 - \jj b_5 &-b_4 + \jj b_8 & b_2 - \jj b_6\\
    b_2 + \jj b_6 & b_4 + \jj b_8 & b_1 + \jj b_5 & b_3 + \jj b_7\\
   -b_4 + \jj b_8 & b_2 - \jj b_6 &-b_3 + \jj b_7 & b_1 - \jj b_5\\
   \end{array}\right. \notag \\
   &\qquad\left.\begin{array}{rrrrr}
    b_1 + \jj b_5 & b_3 + \jj b_7 & b_2 + \jj b_6 & b_4 + \jj b_8  \\
   -b_3 + \jj b_7 & b_1 - \jj b_5 &-b_4 + \jj b_8 & b_2 - \jj b_6  \\
    b_2 + \jj b_6 & b_4 + \jj b_8 & b_1 + \jj b_5 & b_3 + \jj b_7  \\
   -b_4 + \jj b_8 & b_2 - \jj b_6 &-b_3 + \jj b_7 & b_1 - \jj b_5  \\
    a_1 + \jj a_5 & a_3 + \jj a_7 & a_2 + \jj a_6 & a_4 + \jj a_8  \\
   -a_3 + \jj a_7 & a_1 - \jj a_5 &-a_4 + \jj a_8 & a_2 - \jj a_6  \\
    a_2 + \jj a_6 & a_4 + \jj a_8 & a_1 + \jj a_5 & a_3 + \jj a_7  \\
   -a_4 + \jj a_8 & a_2 - \jj a_6 &-a_3 + \jj a_7 & a_1 - \jj a_5
   \end{array}\right].
\end{align}

The code rate of 4Gp-QSTBC for 8 Tx antennas is one symbol pcu. In
general,  by construction, the rate of 4Gp-QSTBC for $2M$ Tx
antennas is the same as the rate of MDC-QSTBC for $M$ Tx antennas.
The maximal rate of MDC-QSTBC is one  symbol pcu \cite{yuen05b}, the
maximal achievable rate of 4Gp-QSTBC is also one  symbol pcu for
$2^m$ Tx antennas. If the number of Tx antennas is $M < 2^m$ $(m = 2, 3, \ldots)$, then
$(2^m - M)$ columns of the code matrix for $2^m$ Tx antennas can be
deleted to obtain the code for $M$ antennas. Thus, \emph{the maximum
rate of 4Gp-QSTBC is one symbol pcu and it is achievable for any
number of Tx antennas}. Additionally, the $4 \times 4$ code matrix
$F_4$ is square. By recursive construction \eqref{e13}, the code matrices of 4Gp-QSTBC are also square for  $2^m$ Tx antennas; and therefore, 4Gp-QSTBC are delay optimal if the number of Tx antennas is $2^m$ \cite{khan06a}.

\subsection{Decoding}

We know that the symbols $s_1, s_2, s_3, s_4$ of $F_4$ can be
separately  detected \cite{yuen05b}. Therefore, from Theorem
\ref{th2}, the 4 groups of 8 symbols of $F_8$ can be detected
independently. These 4 groups are $(s_1, s_2), (s_3, s_4), (s_5,
s_6)$, and $(s_7, s_8)$. The ML metric given in \eqref{e6} can be
derived to detect the 4 groups of symbols of $F_8$. However, to
provide more insights into the decoding of 4Gp-QSTBC, we will derive
an equivalent code and the equivalent channel of $F_8$. Furthermore,
using the equivalent channel of $F_8$, we can use a sphere decoder
\cite{damen03a} to reduce the complexity of the ML search.

The equivalent code of $F_8$ is obtained by column permutations for
the code matrix of $F_8$  in \eqref{e15}: the order of columns is
changed to (1, 3, 5, 7, 2, 4, 6, 8). This order of permutations is
also applied for the rows of $F_8$. Let $x_1 = a_1 + \jj a_5, x_2 =
a_2 + \jj a_6,   x_3 = b_1 + \jj b_5, x_4 = b_2 + \jj b_6, x_5 = a_3
+ \jj a_7, x_6 = a_4 + \jj a_8, x_7 = b_3 + \jj b_7, x_8 = b_4 + \jj
b_8$ be the intermediate variables, we obtain a
permutation-equivalent code of $F_8$ below
\begin{align} \label{e16} %\phantom{-}
    D = \begin{bmatrix*}[r]
        \D_1   & {\D}_2 \\
       -{\D}_2^* & \D_1^* \\
     \end{bmatrix*}
\end{align}
where
\begin{align} \label{e17}
    {\D}_1 = \begin{bmatrix*}[r]
         x_1 & x_2 & x_3 & x_4  \\
         x_2 & x_1 & x_4 & x_3  \\
         x_3 & x_4 & x_1 & x_2 \\
         x_4 & x_3 & x_2 & x_1 \\
     \end{bmatrix*}, \quad
    {\D}_2 = \begin{bmatrix*}[r]
         x_5 & x_6 & x_7 & x_8 \\
         x_6 & x_5 & x_8 & x_7 \\
         x_7 & x_8 & x_5 & x_6  \\
         x_8 & x_7 & x_6 & x_5  \\
     \end{bmatrix*}.
\end{align}
The sub-matrices ${\D}_1$ and ${\D}_2$ have a special form called \emph{block-circulant matrix with circulant blocks} \cite{davis79}.

We next show how to decode the code $D$. For simplicity,  a single
Rx antenna is considered. The generalization for multiple Rx
antennas is straightforward. Assume that the Tx symbols are drawn
from a constellation with unit average power, the Tx-Rx signal model
in \eqref{e3} for the case of STBC $D$ follows
\begin{align} \label{e18}
    \Vect{y} = \sqrt{{\rho}/{8}} D \Vect{h} + \Vect{z}.
\end{align}
Let $\Vect{x} = \begin{bmatrix} x_1 & x_2 & \ldots & x_8 \end{bmatrix} ^{\T}$, $\hat{\Vect{y}} = \begin{bmatrix} y_1 & \ldots & y_4 & y_5^* & \ldots & y_8^* \end{bmatrix} ^{\T}$, $\hat{\Vect{z}} = \begin{bmatrix} z_1 & \ldots & z_4 & z_5^* & \ldots & z_8^* \end{bmatrix} ^{\T}$, and
\begin{align} \label{e19}
    {\HH}_1 = \begin{bmatrix*}[r]
         h_1 & h_2 & h_3 & h_4  \\
         h_2 & h_1 & h_4 & h_3  \\
         h_3 & h_4 & h_1 & h_2 \\
         h_4 & h_3 & h_2 & h_1 \\
     \end{bmatrix*}, \quad
    {\HH}_2 = \begin{bmatrix*}[r]
         h_5 & h_6 & h_7 & h_8 \\
         h_6 & h_5 & h_8 & h_7 \\
         h_7 & h_8 & h_5 & h_6  \\
         h_8 & h_7 & h_6 & h_5  \\
     \end{bmatrix*}.
\end{align}
We have an equivalent expression of \eqref{e18} as
\begin{align} \label{e20}
    \hat{\Vect{y}} = \sqrt{\frac{\rho}{8}} \underbrace{\begin{bmatrix*}[r]
        \HH_1   & \HH_2 \\
        \HH_2^* & -\HH_1^* \\
     \end{bmatrix*}}_{\bar{\HH}} \Vect{x} + \hat{\Vect{z}}.
\end{align}
Note that $\HH_1$ and $\HH_2$ are block-circulant matrices with circulant-blocks  \cite{davis79}. Thus, they are commutative and so do  $\HH_1^*$ and $\HH_2^*$. We can multiply both sides of \eqref{e20} with $\bar{\HH}^{\dag}$ to get
\begin{align} \label{e21}
    \underbrace{\bar{\HH}^{\dag}\hat{\Vect{y}}}_{\bar{\Vect{y}}} = \sqrt{\frac{\rho}{8}} \begin{bmatrix*}
        \HH_1^*\HH_1 + \HH_2^*\HH_2   & \0 \\
        \0  & \HH_1^*\HH_1 + \HH_2^*\HH_2 \\
     \end{bmatrix*} \Vect{x} + \underbrace{\bar{\HH}^{\dag} \hat{\Vect{z}}}_{\bar{\Vect{z}}}.
\end{align}
It can be shown  that the noise  elements of vector $\bar{\Vect{z}}$
are correlated with covariance matrix $\bar{\HH}^{\dag} \bar{\HH}$.
Thus this noise vector can be whitened by multiplying both side of
\eqref{e21} with the matrix $(\bar{\HH}^{\dag} \bar{\HH})^{-1/2}$.
Let $\hat{\HH} = \HH_1^*\HH_1 + \HH_2^*\HH_2$. After the noise
whitening step,  \eqref{e21} is equivalent to the following
equations
\begin{equation} \label{e22}
       \hat{\HH}^{-1/2} \bar{\Vect{y}}_i = \sqrt{\frac{\rho}{8}} \hat{\HH}^{1/2} \Vect{x}_i + \bar{\Vect{z}}_i, \qquad (i = 1, 2),
\end{equation}
where $\bar{\Vect{y}}_i = \begin{bmatrix} \bar{y}_{4i-3} & \bar{y}_{4i-2} & \bar{y}_{4i-1} &  \bar{y}_{4i} \end{bmatrix}^{\T}$,   ${\Vect{x}}_i = \begin{bmatrix} x_{4i-3} & x_{4i-2} & x_{4i-1} &  x_{4i} \end{bmatrix}^{\T}$, the noise vectors\\ $\bar{\Vect{z}}_i = \hat{\HH}^{-1/2} \begin{bmatrix} \bar{z}_{4i-3} & \bar{z}_{4i-2} & \bar{z}_{4i-1} &  \bar{z}_{4i} \end{bmatrix}^{\T}$ are uncorrelated and have elements $\sim \CN(0,1)$.

At this point, the decoding of the 8 transmitted symbols of the code $D$ can be readily decoupled into 2 groups.
However, since the code is a 4Gp-STBC, we can further decompose them
into 4 groups in the following.

Denote the $2\times 2$ (real) discrete Fourier transform (DFT) matrix by $\Ff_2 =
\begin{bmatrix*}[r] 1 & 1\\ 1 & -1
\end{bmatrix*}$. The block-circulant matrices $\HH_1$ and $\HH_2$
can be diagonalized by a (real) unitary matrix $\Theta = \frac{1}{2}
\Ff_2 \otimes \Ff_2$ \cite[Theorem 5.8.2, p. 185]{davis79}. Note
that $\Theta^{\dag} = \Theta$, therefore, ${\HH}_1 = \Theta \Lambda_1 \Theta
\label{e23a}$ and ${\HH}_2 = \Theta \Lambda_2 \Theta \label{e23b}$,
where $\Lambda_1$ and $\Lambda_2$ are diagonal matrices, with eigenvalues of ${\HH}_1$ and  ${\HH}_2$ in the main diagonal, respectively. Thus, $\hat{\HH} = \Theta (\Lambda_1^\dag \Lambda_1 + \Lambda_2^\dag \Lambda_2 ) \Theta$,
and also $\hat{\HH}^{1/2} = \Theta (\Lambda_1^\dag \Lambda_1 +
\Lambda_2^\dag \Lambda_2 )^{1/2} \Theta$.
Since $\hat{\HH}^{1/2}$ is a real matrix, \eqref{e22} becomes
\begin{subequations} \label{e26}
    \begin{eqnarray}
       \hat{\HH}^{-1/2} \Re(\bar{\Vect{y}}_i) = \sqrt{{\rho}/{8}}\hat{\HH}^{1/2} \Re(\Vect{x}_i) + \Re(\bar{\Vect{z}}_i), \qquad i = 1, 2, \label{e26a}\\
       \hat{\HH}^{-1/2} \Im(\bar{\Vect{y}}_i) = \sqrt{{\rho}/{8}}\hat{\HH}^{1/2} \Im(\Vect{x}_i) + \Im(\bar{\Vect{z}}_i), \qquad i = 1, 2. \label{e26b}
    \end{eqnarray}
\end{subequations}
Note that $\Re(\Vect{x}_1) = \begin{bmatrix} a_1 & a_2 & b_1 & b_2 \end{bmatrix}^{\T} := \Vect{d}_1$, i.e. $\Re(\Vect{x}_1)$ is only dependent on the complex symbols $s_1$ and $s_2$. Similarly, $\Re(\Vect{x}_2), \Im(\Vect{x}_1)$, and $\Im(\Vect{x}_2)$ depend on $(s_3, s_4), (s_5, s_6)$, and $(s_7, s_8)$, respectively.

Eq. \eqref{e26} shows that  the decoding of 8 transmitted symbols of
STBC $D$ is separated into  the decoding of 4 groups, each with two
symbols (thus the search space size has been  reduced from $Q^8$ to
$ 4 Q^2$ where $Q$ is the transmit constellation size). A  sphere
decoder \cite{damen03a} can also be used to reduce the complexity of
the ML search for each group. The matrix $\hat{\HH}^{1/2}$ can be
considered as the \emph{equivalent channel} of the 4Gp-QSTBC $D$.

\subsection{Performance Analysis}

In \eqref{e26}, the PEP of the four transmit symbol vectors are the
same. We thus  need to consider the PEP of one of the  vectors
$\Vect{d}_1 = \Re(\Vect{x}_1) =
\begin{bmatrix} a_1 & a_2 & b_1 & b_2 \end{bmatrix}^{\T}$. For
notational simplicity, the subindex $1$ of $\Vect{d}_1$ is dropped.
Additionally, we can introduce redundancy on the signal space by
using a $4\times 4$ real unitary rotation $R$ to the data vector
$\begin{bmatrix} a_1 & a_2 & b_1 & b_2 \end{bmatrix}^{\T}$. Thus the
data vector $\Vect{d} = R \begin{bmatrix} a_1 & a_2 & b_1 & b_2
\end{bmatrix}^{\T}$.

From \eqref{e26a}, the PEP of the pair $\Vect{d}$ and
$\bar{\Vect{d}}$ can be expressed by  the Gaussian tail function as
\cite{simon00b}
\begin{align} \label{e28}
    P(\Vect{d} \rightarrow \bar{\Vect{d}}  |\hat{\HH}) &= Q \left( \sqrt{\frac{\rho}{8} \frac{\|\hat{\HH}^{1/2} R \Vect{\delta} \|^2_{\F}}{ 4 N_0}} \right) \notag \\
    &= Q \left( \sqrt{\frac{{\rho}\left[ \Vect{\delta}^{\T} R^{\T} \Theta^{\T} (\Lambda_1^\dag \Lambda_1 + \Lambda_2^\dag \Lambda_2) \Theta R \Vect{\delta}\right] }{16}} \right).
\end{align}
where $\Vect{\delta} =  \Vect{d} - \bar{\Vect{d}}$, $N_0 = 1/2$ is the variance of  the elements of the white noise vector $\Re(\Vect{z}_1)$ in \eqref{e26a}.

Remember that  $\Lambda_1$ is a diagonal matrix with eigenvalues of $\HH_1$ on the main diagonal. Let $\lambda_{i,j}$ $(i = 1, 2; j =1, 2, 3, 4)$ be the eigenvalues of $\HH_i$. Then $\Lambda_i = \diag\left( \lambda_{i,1}, \lambda_{i,2}, \lambda_{i,3}, \lambda_{i,4} \right)$. Let $\Vect{\beta} = \Theta R \Vect{\delta}$, we have
\begin{align} \label{e30}
    P(\Vect{d} \rightarrow \bar{\Vect{d}}  |\hat{\HH}) = Q \left( \sqrt{\frac{{\rho}( \sum_{i=1}^{2}\sum_{j=1}^{4} \beta_j^2 |\lambda_{i,j}|^2 ) }{16}} \right).
\end{align}

To derive a closed form of \eqref{e30}, we need to evaluate the distribution of $\lambda_{i,j}$. The eigenvectors of  $\HH_1$ is the columns of the matrix $\Theta = \frac{1}{2}{\Ff}_2 \otimes {\Ff}_2$. Thus, the eigenvalues of $\HH_1$ are: $\begin{bmatrix} \lambda_{1,1} & \lambda_{1,2} & \lambda_{1,3} & \lambda_{1,4} \end{bmatrix}^{\T} = ({\Ff}_2 \otimes {\Ff}_2)  \begin{bmatrix} h_1 & h_2 & h_3 & h_4 \end{bmatrix}^{\T}$.
Since $h_j$ $\sim \CN(0,1)$ for  $(j = 1, \ldots , 4)$, thus $\lambda_{1,j}$ $\sim \CN(0,4)$ and so do $\lambda_{2,j}$.

 We now use the Craig's formula \cite{cra91a} to derive the conditional PEP in \eqref{e30}.
\begin{align} \label{e32}
    &P(\Vect{d} \rightarrow \bar{\Vect{d}}  |\hat{\HH}) = Q \left( \sqrt{\frac{{\rho}( \sum_{i=1}^{2}\sum_{j=1}^{4} \beta_j^2 |\lambda_{i,j}|^2 ) }{16}} \right) \notag \\
 &\quad=         \frac{1}{\pi}\int_0^{\pi/2} \exp{\left(\frac{-{\rho}( \sum_{i=1}^{2}\sum_{j=1}^{4} \beta_j^2 |\lambda_{i,j}|^2 )}{32 \sin^2 \alpha }\right) d\alpha }.
\end{align}

Applying a method based on the moment generating function \cite{simon00b},  we obtain the unconditional PEP as: %[Chapters 5 and 9]
\begin{align} \label{e33}
    &P(\Vect{d} \rightarrow \bar{\Vect{d}})
    =  \frac{1}{\pi} \int_0^{\pi/2} \left[\prod_{i=1}^{4}\left(1 + \frac{\rho \beta_i^2}{8\sin^2 \alpha} \right)\right]^{-2} d\alpha.
\end{align}

If $\beta_i \neq 0 \forall i =1, \ldots, 4$, then  $1 + \frac{\rho \beta_i^2}{8\sin^2 \alpha} \approx \frac{\rho \beta_i^2}{8\sin^2 \alpha}$ at high SNR, the approximation of the exact PEP in \eqref{e33} is
\begin{align} \label{e37}
    P(\Vect{d} \rightarrow \bar{\Vect{d}})
    &\approx  \left(\frac{2^{24}{\rho}^{-8} }{\pi} \int_0^{\pi/2} {(\sin\alpha)^{16} d\alpha} \right)
    \prod_{i=1}^{4}|\beta_i|^{-4} \notag \\
    &= \frac{2^7 16! {\rho}^{-8}}{ 8! 8!} \prod_{i=1}^{4}|\beta_i|^{-4}.
\end{align}

The exponent of SNR in \eqref{e37} is -8. This indicates that the
maximum diversity  order of 4Gp-QSTBC is 8 and it is achievable if
the product distance $\prod_{i=1}^{4}\beta_i$ (see \cite{bay04} and
references therein) is nonzero for all possible data vectors.
Furthermore, at high SNR, the asymptotic PEP becomes very tight to
the exact PEP. Recall that $\Vect{\beta} = \Theta R (\Vect{d} -
\bar{\Vect{d}})$; thus, the product matrix $\Theta R$ is the
combined rotation matrix for data vector $\Vect{d}$. Since $\Theta$
is a constant matrix, we can optimize the matrix $R$ so that the
minimum product distance $d_{p,\min} = \min_{\forall \Vect{d}^i,
\Vect{d}^j}  \prod_{k=1}^{4}|\beta_k|$, where $\Vect{\beta} =
\left[\Theta R (\Vect{d}^i - \Vect{d}^j)\right]$ is nonzero and
maximized.

If the complex signals are drawn from QAM, the (real) elements of $\Vect{d}$ are in the set $\{\pm 1, \pm 3, \pm 5, \ldots\}$. The best known rotations for QAM in terms of maximizing the minimum product distance are provided in \cite{bay04, Oggier04}. Denoting the rotation matrix in \cite{bay04, Oggier04} by $R_{BOV}$, the signal rotation for our 4Gp-QSTBC is given by
\begin{align} \label{e39}
    R = \Theta R_{BOV}.
\end{align}
Simulations show that the above vector signal rotation perform
better than the symbol-wise rotation proposed in \cite{yuen05d}
(details  omitted for brevity). We have presented  important
properties of 4Gp-QSTBC. In the next section, we will investigate
4Gp-SAST codes.

\section{Four-Group Decodable STBC Derived from SAST Codes}

\subsection{Encoding}

The SAST code matrix is constructed for $M = 2\bar{M}$ Tx antennas
using circulant blocks.  Two length-$\bar{M}$ data vectors
$\Vect{s}_1 = \begin{bmatrix} s_1 & s_2 & \ldots & s_{\bar{M}}
\end{bmatrix}^{\T}$ and $\Vect{s}_2 = \begin{bmatrix} s_{\bar{M}+1}
& s_{\bar{M}+2} & \ldots & s_{2\bar{M}} \end{bmatrix}^{\T}$ are used
to generate two $\bar{M}$-by-$\bar{M}$ circulant matrices
\cite{davis79}.
Note that the first row of circulant matrix ${\Cc}(\Vect{x})$ copies the row vector  $\Vect{x}$; the $i$th row is obtained by circular shift ($i-1$) times to the right the vector $\Vect{x}$. The SAST code matrix is constructed  as %from ${\Cc}(\Vect{s}_1^{\T})$ and ${\Cc}(\Vect{s}_2^{\T})$
\begin{align} \label{e43}
  \Ss = \begin{bmatrix*}[r]
     \Cc(\Vect{s}_1^{\T}) & \Cc(\Vect{s}_2^{\T}) \\
     -{\Cc}^{\dag}(\Vect{s}_2^{\T}) & {\Cc}^{\dag}(\Vect{s}_1^{\T})
  \end{bmatrix*}.
\end{align}
By construction, 4Gp-SAST codes have rate of one symbol pcu; the
code matrices for an even number of Tx antennas are square; thus
4Gp-SAST codes are delay-optimal for even number of Tx antennas.

\subsection{Decoder of 4Gp-SAST codes}

Similar to 4Gp-QSTBC, the decoding of 4Gp-SAST codes requires two
steps. First,  the two data vectors $\Vect{s}_1$ and $\Vect{s}_2$
are decoupled \cite{dao05a}; then, the real and imaginary parts of
vectors $\Vect{s}_1$ and $\Vect{s}_2$ are separated. We provide the
detail decoder with only one Rx antenna as generalization for multiple Rx
antennas can be easily done.

We introduce another type of circulant matrix called left ciculant,
denoted by $\mathcal{C}_L(\Vect{x})$, where the $i$th row is
obtained by circular shifts ($i-1$) times to the left for the row vector
$\Vect{x}$.

Let us define a permutation $\Pi$ on an arbitrary $M\times M$ matrix $X$ such that,
the $(M-i+2)$th row is permuted with the $i$th row for $i = 2, 3,
..., \big\lceil\frac{M}{2}\big\rceil$, where $\lceil(\cdot)\rceil$
is the ceiling function. One can verify that
\begin{equation}\label{e49}
    \Pi(\mathcal{C}_L(\Vect{x})) = \mathcal{C}(\Vect{x})\,.
\end{equation}

Let $\Vect{y} = \begin{bmatrix} \Vect{y}_1^{\T} & \Vect{y}_2^{\T}\end{bmatrix}^{\T}$, $\Vect{y}_1 = \begin{bmatrix} y_1 & y_2 & \dots & y_{\bar{M}} \end{bmatrix}^{\T}$, $\Vect{y}_2
= \begin{bmatrix} y_{\bar{M}+1} & y_{\bar{M}+2} & \dots & y_{M}\end{bmatrix}^{\T}$, $\Vect{h} = \begin{bmatrix}\Vect{h}_1^{\T} & \Vect{h}_2^{\T}\end{bmatrix}^{\T}$, $\Vect{h}_1 = \begin{bmatrix} h_1 & h_2 & \dots & h_{\bar{M}} \end{bmatrix}^{\T}$, $\Vect{h}_2 = \begin{bmatrix} h_{\bar{M}+1} & h_{\bar{M}+2} & \dots & h_{2\bar{M}}\end{bmatrix}^{\T}$,
$\Vect{z} = \begin{bmatrix}\Vect{z}_1^{\T} & \Vect{z}_2^{\T}\end{bmatrix}^{\T}$, $\Vect{z}_1 = \begin{bmatrix} z_1 & z_2 & \dots &  z_{\bar{M}}\end{bmatrix}^{\T}$, $\Vect{z}_2 =
\begin{bmatrix} z_{\bar{M}+1} & z_{\bar{M}+2} & \dots & z_{2\bar{M}}\end{bmatrix}^{\T}$. We can write the Tx-Rx signal relation as
\begin{equation}\label{e50}
    \begin{bmatrix}
  \Vect{y}_1 \\
  \Vect{y}_2 \\
\end{bmatrix} = \sqrt{\frac{\rho}{M}}\begin{bmatrix*}[r]
     \Cc(\Vect{s}_1) & \Cc(\Vect{s}_2) \\
     -{\Cc}^{\dag}(\Vect{s}_2) & {\Cc}^{\dag}(\Vect{s}_1)
  \end{bmatrix*} \begin{bmatrix}
  \Vect{h}_1 \\
  \Vect{h}_2 \\
\end{bmatrix} + \begin{bmatrix}
  \Vect{z}_1 \\
  \Vect{z}_2 \\
\end{bmatrix}.
\end{equation}
An equivalent form of \eqref{e50} is
\begin{equation}\label{e51}
\begin{bmatrix}
  \Vect{y}_1 \\
  \Vect{y}_2^* \\
\end{bmatrix} =  \sqrt{\frac{\rho}{M}}
\begin{bmatrix}
  X_1 & X_2 \\
  X_3 & X_4 \\
\end{bmatrix} \begin{bmatrix}
  \Vect{s}_1 \\
  \Vect{s}_2 \\
\end{bmatrix} + \begin{bmatrix}
  \Vect{z}_1 \\
  \Vect{z}_2^* \\
\end{bmatrix}
\end{equation}
where $X_1 = \mathcal{C}_L(\Vect{h}_1^{\T}), X_2 = \mathcal{C}_L(\Vect{h}_2^{\T}), X_3 = \mathcal{C}^\dag(\Vect{h}_2^{\T}), X_4 = -\mathcal{C}^\dag(\Vect{h}_1^{\T})$.

Applying permutation $\Pi$ in \eqref{e49} for the
column matrix $\Vect{y}_1$, we obtain
\begin{align}\label{e52}
 \begin{bmatrix}
  \bar{\Vect{y}}_1 \\
  \bar{\Vect{y}}_2 \\
\end{bmatrix} & \triangleq  \begin{bmatrix}
  \Pi(\Vect{y}_1) \\
  \Vect{y}_2^* \\
\end{bmatrix}  \notag \\
  &= \sqrt{\frac{\rho}{M}} \begin{bmatrix}
  \Pi(X_1) & \Pi(X_2) \\
  X_3 & X_4 \\
\end{bmatrix} \begin{bmatrix}
  \Vect{s}_1 \\
  \Vect{s}_2 \\
\end{bmatrix} + \begin{bmatrix}
  \Pi(\Vect{z}_1) \\
  \Vect{z}_2^* \\
\end{bmatrix} \notag \\
  &= \sqrt{\frac{\rho}{M}} \underbrace{\begin{bmatrix*}[r]
  H_1 & H_2 \\
  H_2^\dag & -H_1^\dag \\
\end{bmatrix*}}_{{\HH}} \begin{bmatrix}
  \Vect{s}_1 \\
  \Vect{s}_2 \\
\end{bmatrix} + \begin{bmatrix}
  \bar{\Vect{z}}_1 \\
  \bar{\Vect{z}}_2 \\
\end{bmatrix}
\end{align}
where $H_1 =
\mathcal{C}(\Vect{h}_1^{\T})$, $H_2 =
\mathcal{C}(\Vect{h}_2^{\T})$, $\bar{\Vect{z}}_1 = \Pi(\Vect{z}_1)$,
$\bar{\Vect{z}}_2 = \Vect{z}_2^*$. The elements of $\bar{\Vect{z}}_1$ and $\bar{\Vect{z}}_2$
 are $\sim \mathcal{CN}(0, 1)$, as elements of ${\Vect{z}}_1$ and ${\Vect{z}}_2$.
We now multiply ${\HH}^\dag$ with both sides of \eqref{e52}. Let $\hat{{\HH}} = H_1^\dag H_1
+ H_2^\dag H_2$, we get
\begin{align}\label{e53}
    \begin{bmatrix}
  \hat{\Vect{y}}_1 \\
  \hat{\Vect{y}}_2 \\
\end{bmatrix} &= {\HH}^\dag \begin{bmatrix}
  \bar{\Vect{y}}_1 \\
  \bar{\Vect{y}}_2 \\
\end{bmatrix} = \sqrt{\frac{\rho}{M}} \begin{bmatrix}
   \hat{{\HH}} & \Vect{0}_{\bar{M}} \\
  \Vect{0}_{\bar{M}} & \hat{{\HH}} \\
\end{bmatrix} \begin{bmatrix}
  \Vect{s}_1 \\
  \Vect{s}_2 \\
\end{bmatrix} + {\HH}^\dag \begin{bmatrix}
  \bar{\Vect{z}}_1 \\
  \bar{\Vect{z}}_2 \\
\end{bmatrix} \notag \\
  &= \sqrt{\frac{\rho}{M}} \begin{bmatrix}
   \hat{{\HH}} & \Vect{0}_{\bar{M}} \\
  \Vect{0}_{\bar{M}} & \hat{{\HH}} \\
\end{bmatrix} \begin{bmatrix}
  \Vect{s}_1 \\
  \Vect{s}_2 \\
\end{bmatrix} + \underbrace{\begin{bmatrix}
  \hat{\Vect{z}}_1 \\
  \hat{\Vect{z}}_2 \\
\end{bmatrix}}_{\hat{\Vect{z}}}\,.
\end{align}

The covariance matrix of the additive noise vector $\hat{\Vect{z}}$ is $E[\hat{\Vect{z}} \hat{\Vect{z}}^\dag] = \begin{bmatrix}
   \hat{{\HH}} & \Vect{0}_{\bar{M}} \\
  \Vect{0}_{\bar{M}} & \hat{{\HH}} \\
\end{bmatrix}$.
Therefore, the noise vectors $\hat{\Vect{z}}_1$ and $\hat{\Vect{z}_s}$ are
uncorrelated and have the same covariance matrix $\hat{{\HH}}$. Thus
$\Vect{s}_1$ and $\Vect{s}_2$ can be decoded
separately using $\hat{\Vect{y}}_i = \hat{\HH} \Vect{s}_i +
\hat{\Vect{z}_i}$, $i = 1, 2$. The noise vectors $\hat{\Vect{z}}_1$ and $\hat{\Vect{z}_s}$ can be whitened by the same whitening matrix $\hat{\HH}^{-1/2}$. The equivalent equations for Tx-Rx signals are
\begin{equation}\label{e55}
        \hat{\HH}^{-1/2}\hat{\Vect{y}}_i = \sqrt{{\rho}/{M}} \hat{\HH}^{1/2} \Vect{s}_i + \hat{\HH}^{-1/2}\hat{\Vect{z}_i}, \qquad i = 1, 2.
\end{equation}

At this point, the decoding of SAST codes becomes the detection of 2 group of complex symbols $\Vect{s}_i$ $(i = 1, 2)$; this is similar to the detection of 4Gp-QSTBC in \eqref{e22}. Our next step is to separate the real and imaginary parts of vectors $\Vect{s}_i$ to obtain 4 groups of symbols for data detection.

Recall that $\hat{\HH} = H_1^\dag H_1 + H_2^\dag H_2$, and both
$H_1$ and $H_2$ are circulant. Hence, $\hat{\HH}$ is also circulant
\cite{davis79}. Let $\Lambda_i = \begin{bmatrix} \lambda_{i,1} &
\lambda_{i,2} & \ldots & \lambda_{i,m} \end{bmatrix}$ be the $m$
eigenvalues of $H_i$ $(i = 1, 2)$. We can diagonalize $H_i$ by
DFT matrix as $H_i = \Ff^\dag \Lambda_i \Ff$. Thus $\hat{\HH} = {\Ff}^\dag (\Lambda_1^\dag \Lambda_1 + \Lambda_2^\dag \Lambda_2) \Ff$.
Let $\Lambda_1^\dag \Lambda_1 + \Lambda_2^\dag \Lambda_2 = \Lambda$,
then $\Lambda$ has real and non-negative entries in the main diagonal and
$\hat{\HH}^{1/2} = {\Ff}^\dag \Lambda^{1/2} \Ff$ and
$\hat{\HH}^{-1/2} = {\Ff}^\dag \Lambda^{-1/2} \Ff$.

We assume that $\Vect{s}_i$ is pre-multiplied (or rotated) by an
IDFT matrix $\Ff^\dag$ of proper size. Substituting $\Vect{s}_i$ by
$\Ff^\dag \Vect{s}_i$ and multiplying both sides of \eqref{e55} with
the DFT matrix $\Ff$, we obtain
\begin{align}\label{e58}
        \Lambda^{-1/2} \Ff \hat{\Vect{y}}_i &= \sqrt{{\rho}/{M}}\Ff \hat{\HH}^{1/2} {\Ff}^\dag \Vect{s}_i + \Lambda^{-1/2} \Ff \hat{\Vect{z}_i}  \notag \\
        &= \sqrt{{\rho}/{M}} \Lambda^{1/2} \Vect{s}_i + \underbrace{\Lambda^{-1/2} \Ff \hat{\Vect{z}_i}}_{\check{\Vect{z}}_i}.
\end{align}
Since $\Lambda^{1/2}$ is a real matrix, the real and imaginary parts of $\Vect{s}_i$ $(i = 1, 2)$ can now be separated for detection.
\begin{subequations} \label{e59}
\begin{align}
        \Lambda^{-1/2} \Re(\Ff \hat{\Vect{y}}_i) &= \sqrt{{\rho}/{M}} \Lambda^{1/2} \Re(\Vect{s}_i) + \Re(\check{\Vect{z}}_i), \label{e59a} \\
        \Lambda^{-1/2} \Im(\Ff \hat{\Vect{y}}_i) &= \sqrt{{\rho}/{M}} \Lambda^{1/2} \Im(\Vect{s}_i) + \Im(\check{\Vect{z}}_i). \label{e59b}
        \end{align}
\end{subequations}

We finish deriving the general decoder for 4Gp-SAST codes. Using \eqref{e59}, one can use a sphere decoder to detect the transmitted symbols. The \emph{equivalent channel} of 4Gp-SAST codes is $\Lambda^{1/2}$.

\subsection{Performance Analysis}

Note that the eigenvalues  of $m\times m$ matrices $H_1$ and $H_2$
can be found easily using unnormalized DFT of the
channel vectors $\Vect{h}_1$ and $\Vect{h}_2$ \cite{davis79}.
Therefore, the eigenvalues of $H_1$ and $H_2$ have distribution
$\sim \CN(0, m)$.

Similar to the case of 4Gp-QSTBC, we can introduce  a real
orthogonal transformation $R$ to the data vectors $\Re(\Vect{s}_i)$
and $\Im(\Vect{s}_i)$ $(i = 1, 2)$ to improve the performance of
4Gp-SAST codes. Thus the actual signal rotation of 4Gp-SAST codes is
${\Ff}^\dag R$.

Since the PEP of vectors $\Re(\Vect{s}_i)$ and $\Im(\Vect{s}_i)$ $(i
= 1, 2)$ are the same, we only calculate the PEP of the vector
$\Re(\Vect{s}_1)$. Let $\Vect{d} = \Re(\Vect{s}_1)$. The PEP of
distinct vectors $\Vect{d}$ and $\bar{\Vect{d}}$ can be calculated
in a similar manner to that of 4Gp-QSTBC in Section III-C.
details are omitted for brevity. The PEP of 4Gp-SAST codes is given
below.
\begin{align} \label{e60}
    P(\Vect{d} \rightarrow \bar{\Vect{d}})
    &=  \frac{1}{\pi} \int_0^{\pi/2} \left[\prod_{i=1}^{m}\left(1 + \frac{\rho \beta_i^2}{8\sin^2 \alpha} \right)\right]^{-2}
    d\alpha
\end{align}
where $\begin{bmatrix} \beta_1 & \beta_2 & \ldots & \beta_m \end{bmatrix}^{\T} = R (\Vect{d} - \bar{\Vect{d}})$. One can find the asymptotic PEP of 4Gp-SAST codes at high SNR in a similar fashion to the case of 4Gp-QSTBC in \eqref{e37} as follows.
\begin{align} \label{e61}
    P(\Vect{d} \rightarrow \bar{\Vect{d}})
    &\approx \left(\frac{2^{6m}{\rho}^{-2m} }{\pi} \int_0^{\pi/2} {(\sin\alpha)^{16} d\alpha} \right) \prod_{i=1}^{m}\beta_i^{-4} \notag \\
    &= \frac{2^{6m}{\rho}^{-2m}}{2^{17}} \frac{16!}{8! 8!} \prod_{i=1}^{m}\beta_i^{-4}.
\end{align}

Thus, if the product distance $\prod_{i=1}^{m}\beta_i$ is nonzero,
4Gp-SAST codes will achieve full-diversity. Similar to 4Gp-QSTBC,
with QAM, the signal rotations $R_{BOV}$ in \cite{bay04, Oggier04}
can be used to minimize the worst-case PEP.

\emph{Remark}: It is interesting to recognize that, the optimal
rotation  matrices of 4Gp-QSTBC ($R = \Theta R_{BOV}$) and 4Gp-SAST codes  ($R = \Ff R_{BOV}$) have a similar formula. The
precoding matrices $\Theta$ and $\Ff$ are added to diagonalize the
channels of the two codes. Thus each real symbol is equivalently
transmitted in a separate channel, but full diversity is not
achievable. The real rotation matrix $R_{BOV}$ is applied to the
data vectors so that the real symbols are spread over all the
channels, and thus full diversity is achievable.

\section{Simulation Results}

\begin{figure}[th]%[bh]
\centering
\includegraphics[width=3.5in]{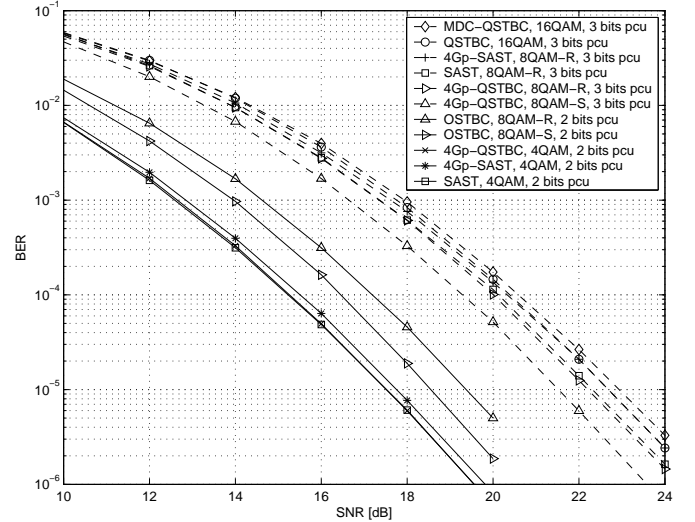}%
\caption{Performances of 4Gp-QSTBC and 4Gp-SAST codes compared with OSTBC, MDC-QSTBC, QSTBC and SAST codes, 6 Tx and 1 Rx antennas, 2 and 3 bits pcu.} \label{f1}
\end{figure}

Simulation results are presented  in Fig. \ref{f1} to compare the
performances of 4Gp-QSTBC and 4Gp-SAST codes  with OSTBC, MDC-QSTBC
\cite{yuen05b}, QSTBC \cite{su04b}, and SAST codes \cite{dao05a} for
6 Tx and 1 Rx antennas. To produce the desired bit rates, two 8QAM
constellations are used. The first constellation is rectangular,
denoted by 8QAM-R, and has signal points $\{\pm 1 \pm \jj, \pm 3 \pm
\jj\}$. The other constellation, denoted by 8QAM-S, has the best
minimum Euclidean distance; its geometrical shape is depicted in
\cite[Fig. 2(c)]{su04b}.

We compare the performance of our new codes with OSTBC and SAST codes for a spectral efficiency of 2 bits pcu. To get this bit rate, 8QAM signals are combined with rate-2/3 OSTBC, while 4QAM is used for the SAST, 4Gp-QSTBC and 4Gp-SAST codes. Two columns (4 and 8) of 4Gp-QSTBC
for 8 Tx antennas is deleted to create the code for 6 Tx
antennas. From Fig. \ref{f1}, 4Gp-SAST codes
gains 0.8 and 1.6 dB over OSTBC with 8QAM-S and 8QAM-R,
respectively, while the decoding complexity slightly increases
(see Table \ref{t1}). The performance improvement of
4Gp-QSTBC is even better, 1 dB compared with OSTBC (using 8QAM-S)
and 0.2 dB compared with 4Gp-SAST codes. Note that for 6 antennas, the decoding
complexity of 4Gp-QSTBC is slightly
higher than that of 4Gp-SAST codes (see Table \ref{t1}).

In Fig. \ref{f1}, the performance of 4Gp-QSTBC and 4Gp-SAST codes with 3
bits pcu is also compared with that of the rate-3/4 QSTBC
and MDC-QSTBC (using 16QAM). 4Gp-SAST code yields a 0.3 dB improvement
over MDC-QSTBC and performs the same as
QSTBC. Specifically, 4Gp-QSTBC using 8QAM-S performs much better than the QSTBC; it produces a 1.2 dB gain over QSTBC with the same decoding complexity.

Further simulations for 5 and 8 Tx antennas also confirm that 4Gp-QSTBC and 4Gp-SAST codes perform better than  OSTBC, MDC-QSTBC, QSTBC, and SAST codes. Due to the lack of space, we omit the details.

\section{Conclusions}

We have presented two new rate-one STBC with four-group decoding,
called 4Gp-QSTBC and 4Gp-SAST codes. They offer the lowest decoding complexity compared with the existing rate-one STBC. Their closed-form PEP are derived, enabling the optimization of signal rotations.  Compared with other existing low decoding
complexity STBC (such as OSTBC, MDC-QSTBC, CIOD, and QSTBC), our
newly designed STBC have several additional advantages including higher code
rate, better BER performance, lower encoding/decoding delay, and lower
peak-to-average power ratio (PAPR) because zero-amplitude symbols
are avoided in the code matrices. Recent results in \cite{kar06a} present a flexible design of multi-group STBC. However, the code rate is still limited by 1 symbol pcu. Thus, the systematic design of high-rate multi-group STBC is still an open research problem.

\section*{Acknowledgment}

The authors would like to thank anonymous reviewers for their constructive comments, which greatly improve the presentation of the paper.

\end{document}